# Relative sea level and coastal vertical movements in relation with volcano-tectonic processes in Mayotte

**Julien Gargani[1, 2]**

[1]Université Paris-Saclay, CNRS, Geops, Orsay, France.

[2]Université Paris-Saclay, Centre d'Alembert, Orsay, France.

**Highlights**

- An abrupt subsidence of 10 m occurred in Mayotte at 9.3 kyr B.P., followed by an uplift of 10 m from 8.1 kyr to 7 kyr B.P.
- Volcano-tectonic activity is responsible for these vertical movements.
- Deep reservoir overpressure has been recorded over a period of 1 kyr.

**Abstract**

During the last 10 kyr, significant subsidence and uplift occurred on Mayotte Island in the Comoros archipelago, but the role of volcanic processes in Holocene vertical movements has been neglected. Here, we show that an abrupt subsidence of 10 m occurred 9.3 kyr ago and then an uplift of the same amplitude occurred at a rate of 9 mm/yr from 8.1 kyr to 7 kyr ago, and we compare the relative sea level of Mayotte with a reference sea level curve. Using a modeling approach, this study shows that an increase and a decrease in pressure in a deep magma reservoir was responsible for the ~10 m subsidence and the 10 m uplift, whereas the loading by new volcanic edifices has caused the subsidence that has occurred during the past 6.1 kyr. The surface movement was caused by pulses from the deep mantle that were probably related to superplume activity.

## 1. Introduction

Recent volcano-seismic events in unexpected places or with significant impacts challenge our understanding of volcanic activity. First, these events suggest reconsidering the role of volcanic activity or sea-level rise/inundation on local natural hazards [Roger, 2019: Audru et al., 2010] and promoting specific studies on the reduction in risk [Deves et al., 2022; Pasquon et al., 2022] associated with volcanic activity. Second, the potential interrelation between short-term volcanic activity and long-term volcano-tectonic evolution also makes it challenging to understand volcanic processes and Earth surface evolution.

The risk associated with volcanic activities is not only due to magmatic explosions and lava flows but also to a consequence of induced landslides [Ward and Day, 2001; Gargani, 2020] and subsidence [Lambeck, 1981], thereby increasing marine submersion risk and tsunamis [Ward and Day, 2001]. Subsidence in a volcanic context may be associated with volcanic edifice loading [Lambeck, 1981; Gargani, 2022b] or magma reservoir deflation [Gargani, 2006]. The relationship between volcanic activity and significant relative sea level variation has been highlighted in previous studies on contemporary observations as well as on past events [Sternai et al., 2017; Duggen et al., 2003].

Mayotte Island (Comoros archipelago, SW Indian Ocean, between Africa and Madagascar) has experienced significant seismic activity since May 2018, with more than 11000 earthquakes of up to magnitude 5.9, unusually long-period events (more than 400) and surface deflation up to 200 mm/yr [Lemoine et al., 2020; Cesca et al., 2021]. This seismic activity indicates deep magma reservoirs at 20-35 km, 50 km and 70 km from Mayotte's Petite Terre Island [Lemoine et al., 2020; Cesca et al., 2021; Feuillet et al., 2021]. This latter reservoir is connected to the 50 km-depth reservoir through rapid magma propagation in the lithosphere [Lavassyere et al., 2022]. The magma reservoir diameters are 10-15 km at depths of 20-35 km and 20-30 km at depths of ~50 km [Lemoine et al., 2020; Cesca et al., 2021; Feuillet et al., 2021; Lavassyere et al., 2022] (Figure 1). During the 2018-2021 seismo-volcanic event, a deflation of the magma reservoir located at a depth of ~50 km occurred (Lavassyere et al., 2022). This seismo-volcanic activity caused the birth of an 820 m tall, 5 km$^3$ deep sea volcanic edifice documented on the eastern insular slopes of Mayotte at the tip of a 50 km-long volcanic ridge composed of many other recent edifices and lava flows [Feuillet et al., 2021].

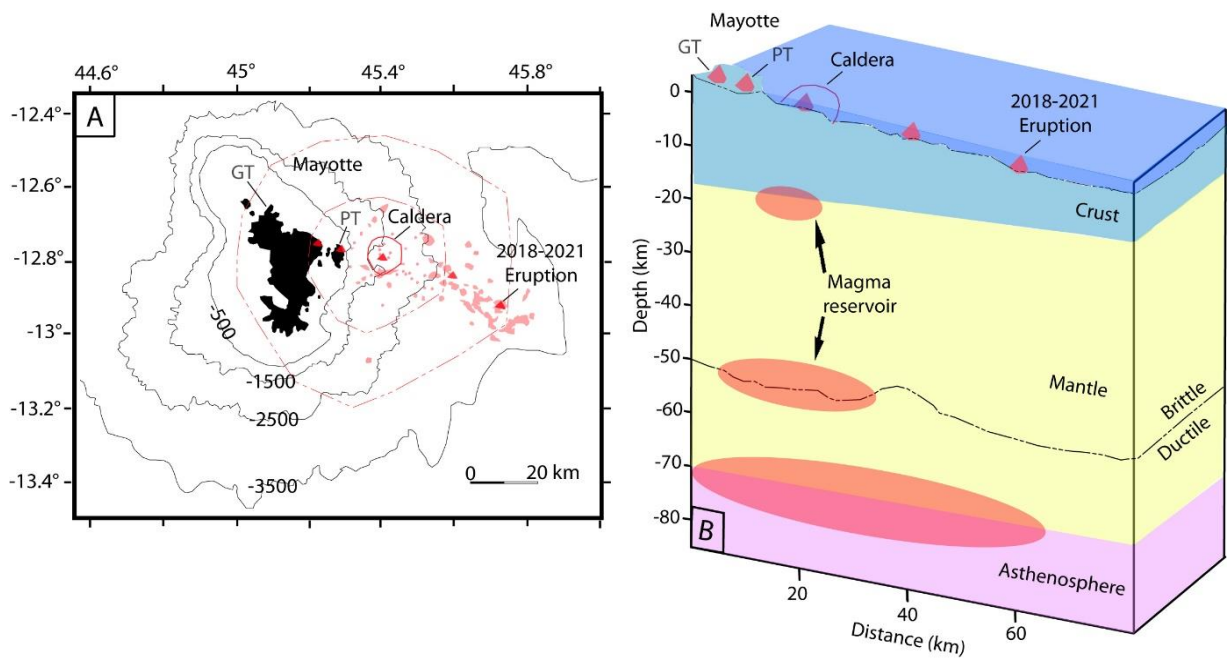

*Figure 1: Mayotte volcanic context. (A) Grand Terre (GT) and Petite Terre (PT) islands in Mayotte archipelago, with a volcanic edifice, lava flows and a caldera, (B) schematic representation of the lithosphere below Mayotte.*

Before the 2018-2021 event, the most recent volcanic activity in the archipelago occurred on the Comoros Islands (Karthala volcano), 200 km away from Mayotte [Bachèlery et al., 2016]. Studies have suggested that the presence of a hotspot explains the E–W sublinear trend of the Comoros volcanic archipelago [Emerick and Duncan, 1982; Ebinger and Sleep, 1998; Class et al., 2005], but this explanation is controversial [Bertil et al., 2021; Quidelleur et al., 2022]. Researchers have interpreted the E–W seismic alignment in the region as highlighting the activity of structures associated with the continuity of the East African Rift and in relation to the Madagascar drift on an E–W-trending axis [Bertil et al., 2021].

The main volcanic building phase of Mayotte Island is estimated to have occurred from 10.6 Ma to 1.25 Ma. The lava content demonstrates lithospheric but also mantle sources, suggesting lithospheric melt as well as the existence of an ascending mantle plume [Pelleter et al., 2014]. The mantle depths range from 17 km to 70 km below Mayotte [Feuillet et al., 2021]. On Mayotte Island, volcanic events have been dated to 6.5-7 kyr B.P. (ash), 8 and 2.2 kyr B.P. (phonolite), 4 kyr and 4.12 ± 0.04 kyr B.P. (pumice stone), and 5.98 ± 0.14 kyr B.P. (volcanic mud) [Zinke et al., 2003a, 2000; Camoin et al., 2004]. These volcanic activities highlight the existence of volcanic edifice building during the Holocene. Nevertheless, the volume of volcanic rock that erupted during this phase is unknown. Furthermore, the duration of the complete volcanic cycle from magma ascent to volcanic edifice construction in Mayotte during the Holocene has never been estimated.

Relative sea level is useful to quantify vertical movements [Gargani, 2022a and 2022b]. Previous studies on coral reefs have obtained accurate data on relative sea levels in Mayotte during the Holocene [Zinke et al. 2001; Zinke et al., 2003b; Zinke et al., 2003a; Zinke et al., 2005; Camoin et al., 1997; Camoin et al., 2004]. These studies have discussed the role of climate variation and postglacial floods, as well as glacio-hydro-isostatic adjustment, on relative sea level. However, the potential effect of volcanic activity on relative sea level rise and fall has not been analyzed in Mayotte.

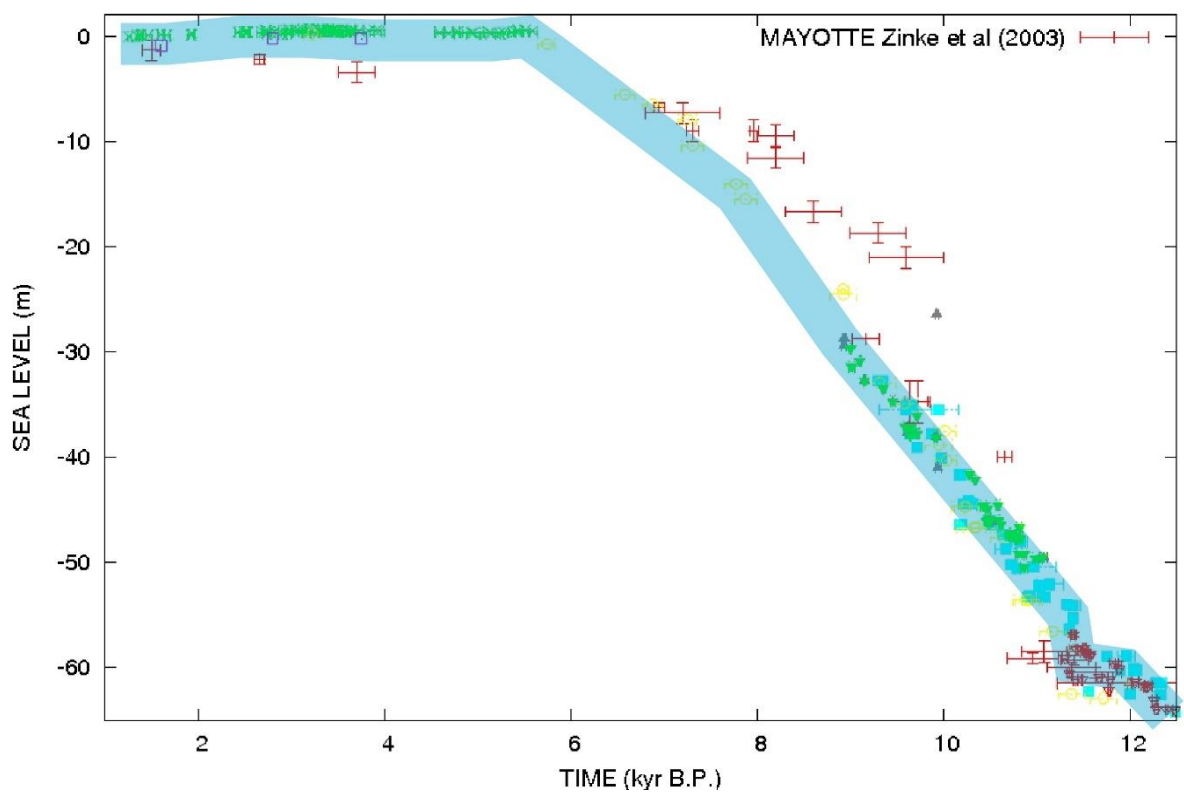

*FIGURE 2: Relative sea level in Mayotte (red cross) compared to the sea level curve constructed using data from Tahiti and Barbados (light blue). Data from Mayotte: Zinke et al. (2003); data from Tahiti: Bar et al. (1990, 1996, 2010); Hallman et al. (2018); Deschamps et al. (2012); Pirazzoli et al. (1985); data from Barbados: Peltier et al. (2006); Abdul et al. (2016). Before 6 kyr B.P., the subsidence rate in Tahiti is corrected with an uplift rate of 0.25*

mm/yr, and afterward, it is corrected using a subsidence rate of 0.15 mm/yr. Before 11.18 kyr, the uplift rate is corrected with an uplift rate of 0.34 mm/yr in Barbados, and after 11.18 kyr B.P., an uplift rate of 0.8 mm/yr is used. $^{14}C$ ages are corrected by a U/Th ratio of 1.14.

This study analyzes the vertical movement in Mayotte during the last 10 kyr by modeling magma reservoir pressure as well as volcanic edifice loading and compares these parameters with relative sea level. More precisely, volcanic activities, such as magma reservoir overpressure (amplitude and duration) and the volume of volcanic edifice during the last 10 kyr in Mayotte, are quantified. The implications of these results at the regional scale are discussed.

## 2. Method

### 2.1 Overpressure and surface uplift

The vertical displacement $U_z$ of a magma reservoir of radius $r_s$ at a depth $-Z$ with an overpressure $\Delta P$ is given by the following equation [Battaglia et al., 2013; McTigue, 1987]:

$U_z = (1-\nu) (\Delta P\, r_s^3\, Z) / [G\, (R^2+Z^2)]^{3/2} \times [1 - (r_s/Z)^3 \times [(1-\nu) / [2\,(7-5\nu)] - [15\,(2-\nu)] / [4\,(7-5\nu)] \times Z^2/(R^2+Z^2)\,]]$

where $\nu$ is the Poisson's ratio ($\nu=0.25$) and R is the distance from the expansion source embedded at position $\xi$ to point $j$ on the free surface and corresponds to the radial distance from the source. G is the shear modulus (15 GPa).

The depths of the magma reservoir that are considered in this study are 20 km, 50 km and 70 km, which are in agreement with the estimation by Feuillet et al. (2021). The radii of the magma reservoir tested by modeling range from 5 km to 30 km. The influence of the magma reservoir pressure at different depths was also modeled. The overpressure and depressurization tested in this study range from 20 to 150 MPa. The overpressure was estimated using the observed uplift of $10 \pm 1$ m.

### 2.2 Volcanic edifice loading and isostatic adjustment

Loading of volcanic edifices on the lithosphere causes subsidence [Lambeck, 1981]. Modeling isostatic adjustment in response to loading is performed using a classical law [Turcotte and Schubert, 2001] that has been applied in various studies [Watts, 2000; Gargani, 2004 et 2010]. Turcotte and Schubert (2001) used the equation $\nabla^2(D \cdot \nabla^2 w(x)) + (\rho_a+\rho_v)\, g \cdot w(x) = \rho_v g\,[z_{init}(x) + z(x)]$ to model the flexure of the lithosphere, where D is the rigidity of the lithosphere, w(x) represents the uplift, $\rho_a$ and $\rho_v$ are the densities of the asthenosphere and volcanic rock, $g$ is the acceleration of gravity, $z_{init}(x)$ is the initial topography and $z(x)$ is the topography after the construction of the new volcanic edifice. The rigidity is defined by $D = ET_e^3/[12(1-\nu^2)]$, where $E$ is Young's modulus and $T_e$ is the effective elastic thickness.

The E–W sublinear trend of the Comoros volcanic archipelago permits us to analyze the symmetry axis of this 200 km-long structure and to simplify the numerical problem using a 2-D approach. The elastic thickness in Mayotte is $40 \pm 5$ km [Calmant et al., 1990], which corresponds to a flexural rigidity that ranges from $3.8 \times 10^{22}$ N.m$^2$ to $8.1 \times 10^{22}$ N.m$^2$. In the calculation, $E = 10^{10}$ Pa.

The resulting vertical motion rates are calculated by considering a constant displacement during the 10 kyr after the abrupt mass displacement, which is in agreement with the time necessary to relax the viscous properties of the lithosphere $10 \pm 5$ kyr after isostatic adjustment [Mitrovica et al., 2000; Van der Wal et al., 2010]. The average velocity that is obtained is a first-order estimation sufficient to describe the main dynamics of the system.

Different loading masses corresponding to various volumes of volcanic rock involved in volcanic edifice building were simulated. Loading was simulated by several volcanic edifices that were 200 m tall and had a diameter of 3.5 km, corresponding to a conical volume of 0.6 km$^3$ and a mass of 1.8 × 10$^{12}$ kg. This study discusses the loading mass that explains the observed subsidence. Two different locations have been considered (0 km and 20 km from the coast). The results have been compared with observed vertical movements.

*2.3 Relative sea level and vertical movement*

Well-preserved coral reefs were collected in Mayotte and dated [Camoin et al., 1997; Zinke et al., 2004a; Camoin et al., 2004]. A relative sea level curve for Mayotte was obtained [Zinke et al., 2003a; Camoin et al., 2004]. Comparing the relative sea level curve of Mayotte with a reference sea level curve permits us to estimate the vertical movement in Mayotte from 10 kyr to 1 kyr B.P.

The reference sea level curve used in this study is based on Bard et al. [1990; 1996; 2010], Deschamps et al. [2012] and Hallman et al. [2018] for Tahiti and Peltier et al. [2006] and Abdul et al. [2016] for Barbados. For accuracy, only U–Th ages are used for the reference sea level curve. For coherency, only the coral of the Acropora palmata species are included in the reference curve. For the construction of the sea level curve using Tahiti data, subsidence rate corrections of 0.15 mm/yr after 6 kyr B.P. [Deschamps et al., 2018] and of 0.25 mm/yr before 6 kyr B.P. [Bard et al., 1990] are used. For Barbados, uplift rates of 0.34 mm/yr before 11.18 kyr B.P. and 0.8 mm/yr before 11.18 kyr B.P. are used [Bard et al., 2016; Gargani, 2022].

## 3. Results and interpretation

A comparison between the relative sea level in Mayotte and the reference sea level suggests that a subsidence of approximately 10 ± 1 m occurred abruptly at 9.3 kyr B.P. in Mayotte in less than 200 yr at a minimum subsidence rate of 50 mm/yr (Figure 3). Then, an uplift of +10 m occurred from 8.1 to 7 kyr B.P. during 1 kyr at a rate of 9 mm/yr. Afterward, progressive subsidence occurred from 6.1 kyr B.P. to 1 kyr B.P. at a rate of 0.4 mm/yr (i.e., 2 m in 5 kyr).

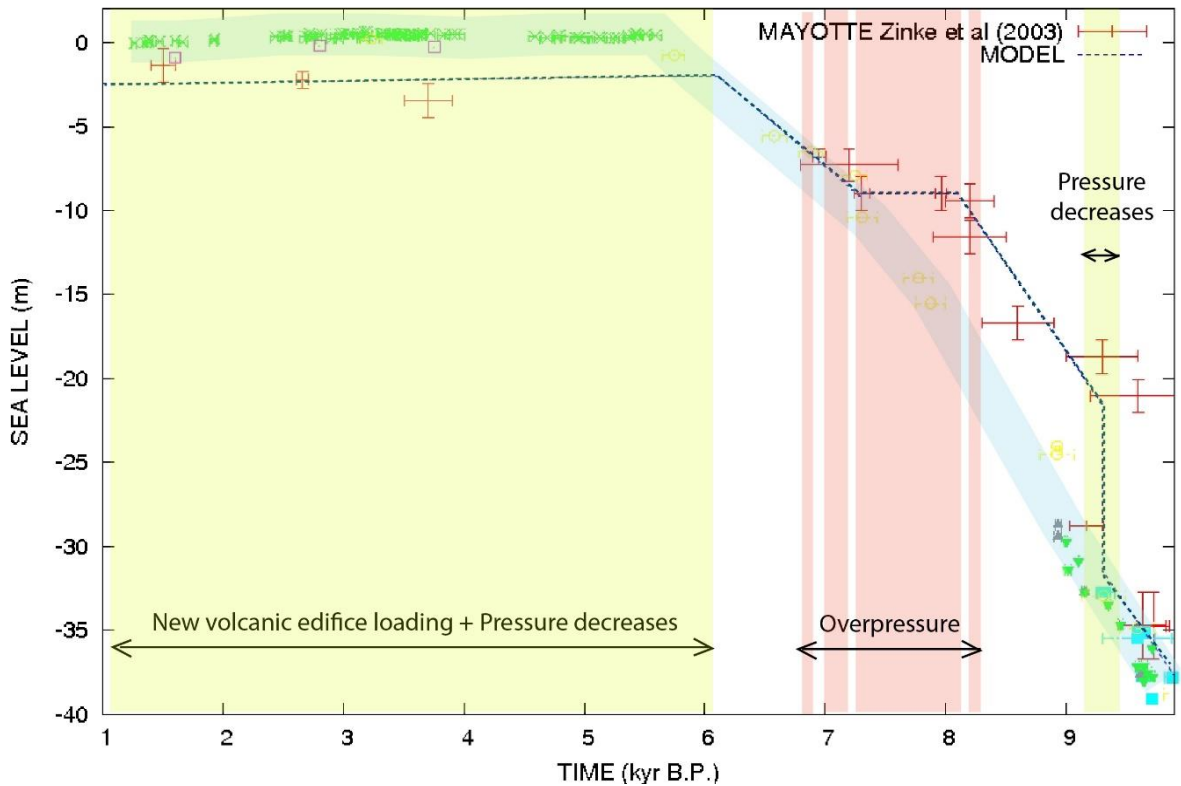

*Figure 3: Comparing the relative sea level variation in Mayotte with a reconstructed sea level curve (light blue curve; see the methods for details) using a modeling approach. Abrupt depressurization of a magma reservoir occurred at 9.3 kyr B.P. and caused a subsidence of ~10 m. Then, a progressive pressure increase in the magma reservoir occurred from 8.1 to 7 kyr B.P. and caused an uplift rate of 9 mm/yr. At 6.1 kyr B.P., new volcanic edifices loaded the lithosphere.*

An increase in the pressure in the magma reservoir associated with magma ascent can explain the uplift of the surface observed between 8.1 and 7 kyr ago. More precisely, a magma reservoir overpressure of 80 ± 20 MPa at a depth of 50 km with a size of 20 km could explain an uplift of 10 ± 1 m (Figure 4). Nevertheless, an overpressure of 80 ± 20 MPa at a depth of 70 km for a magma reservoir with a radius of 25 km could also explain this uplift (Fig. 5).

The cause of the subsidence of 10 m that occurred 9.3 kyr ago in Mayotte is well explained by a decrease in the magma reservoir pressure by the same amount (i.e., a pressure decrease of 80 ± 20 MPa at a depth of 50 km with a magma reservoir size of 20 km, or a pressure decrease of 80 ± 20 MPa at a depth of 70 km for a magma reservoir with a radius of 25 km). With a maximum duration of 200 yr for abrupt subsidence, the minimum subsidence rate was 5 cm/yr.

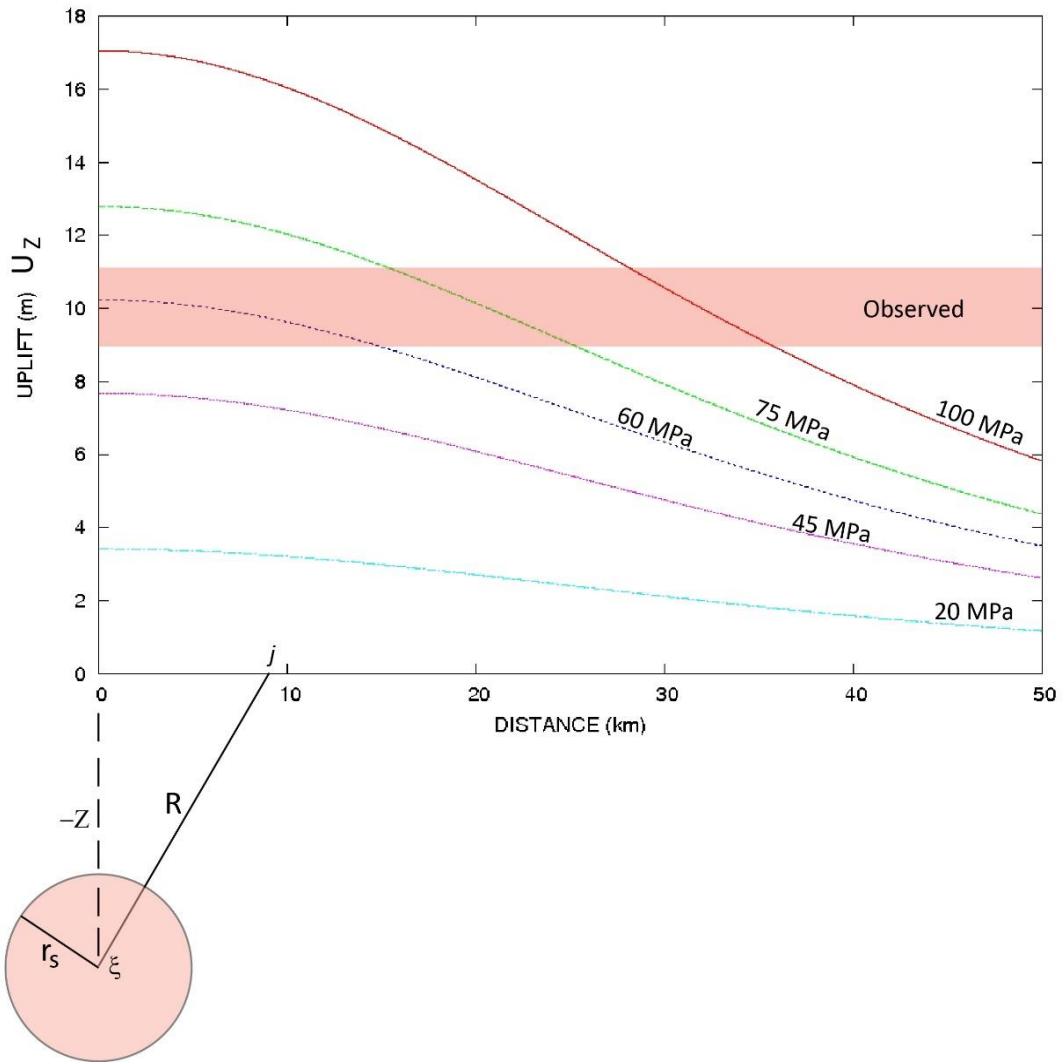

FIGURE 4: Modeling of the vertical displacement $U_z$ caused by magma reservoirs with various overpressures $\Delta P$. The magma reservoir (not to scale) has a radius $r_s$ and is located at a depth $-Z$ at a distance $R$ from the expansion source embedded at position $\xi$ to point $j$. The shear modulus $G=15$ GPa, $\nu=0.25$, $r_s=50$ km, and $Z=50$ km. The observed uplift is $10 \pm 1$ m.

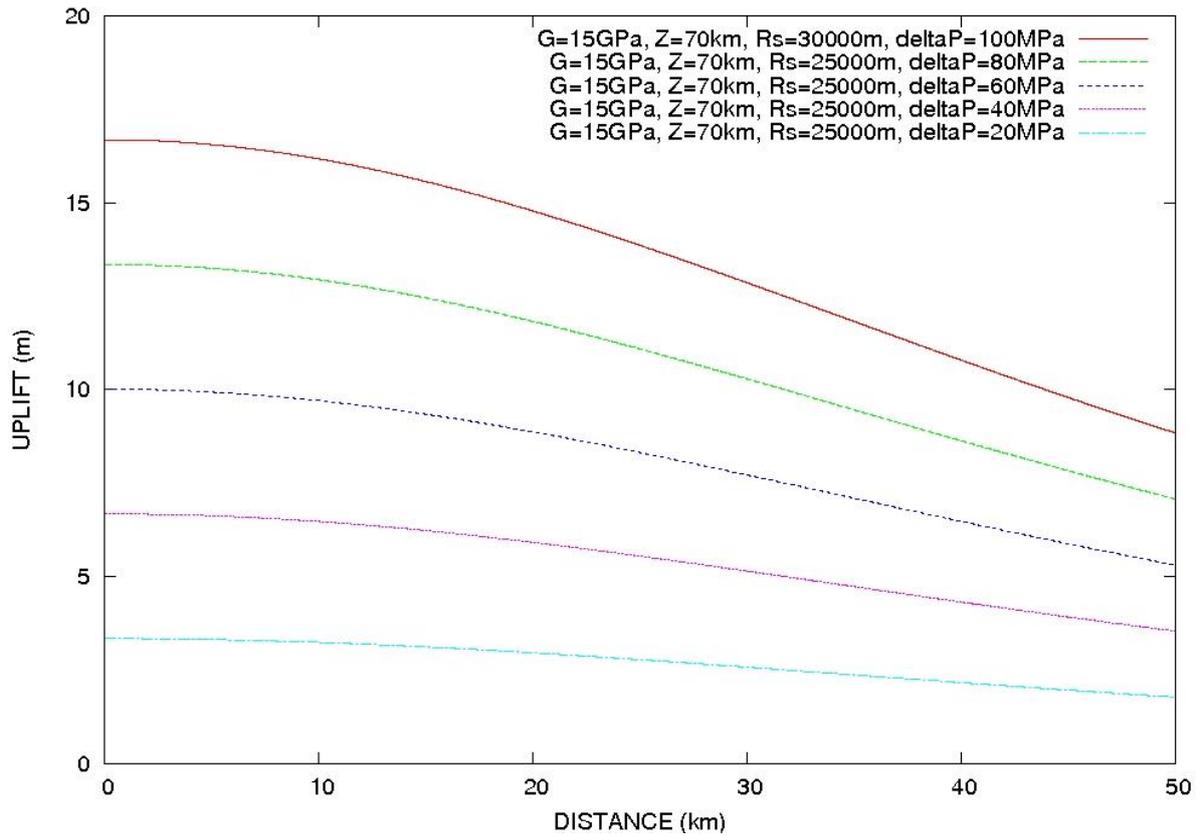

*Figure 5: Modeling the vertical displacement $U_z$ caused by magma reservoir overpressure $\Delta P$ at a depth Z of 70 km with a radius $r_s$ of 25 km. The shear modulus G=15 GPa, and v=0.25. The observed uplift is 10 ± 1 m.*

The subsidence that occurred from 6.1 kyr B.P. to 1 kyr B.P. is due to a progressive deflation of the magma reservoir and/or a loading of the lithosphere by one or more new volcanic edifices. The loading by a volcanic edifice located at a distance of 0 km from Mayotte with a height of 200 m caused a subsidence of 1.4 ± 0.2 m (Figure 6). This value corresponds to a subsidence rate of 0.14 mm/yr, assuming a duration of 10 kyr for isostatic adjustment. Considering a second edifice construction of a height of 200 m at a maximum distance of 40 km from Mayotte (GT), the total subsidence could reach 2.5 m. To obtain the observed subsidence rate of 0.4 mm/yr, it is necessary to load the lithosphere by a mass of 5.4 × $10^{12}$ kg, corresponding to a volume of 1.8 km$^3$ or a deflation of the magma reservoir of 30-40 MPa for a chamber at depths of 50-70 km.

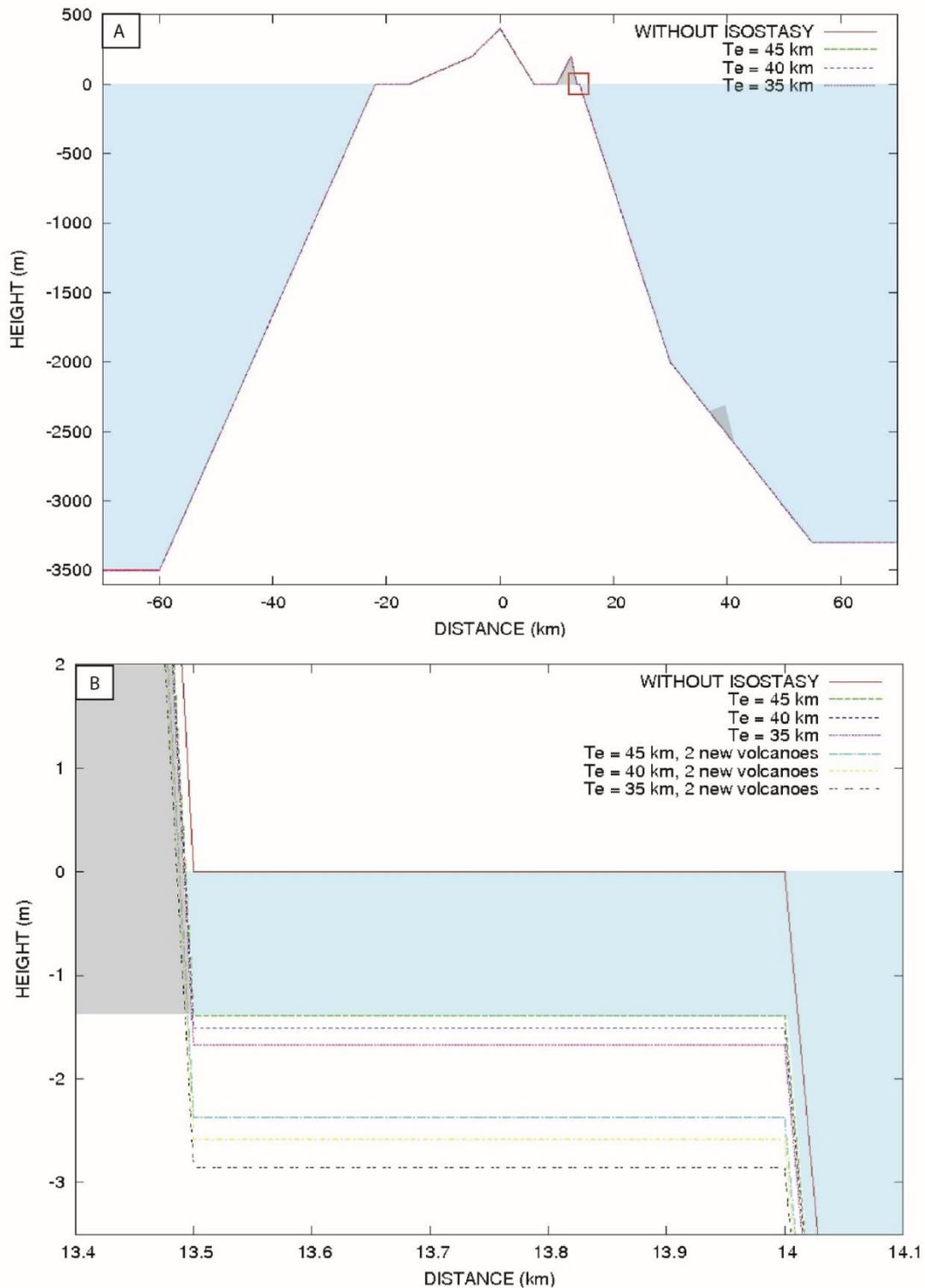

*Figure 6: Isostatic adjustment modeling. (A) Isostatic adjustment modeling for different elastic thicknesses after loading by volcanic edifices located in Petite Terre and the offshore Mayotte archipelago are in gray. (B) Details of the modeling results. ν=0.25, $ρ_a$ =3179 kg/m³, $ρ_v$ = 3000 kg/m³, E = $10^{10}$ Pa, and 35 km < $T_e$ < 45 km.*

## 4. Discussion

The complete vertical movement sequence is characterized by an abrupt subsidence phase of approximately 10 m due to magma chamber deflation at a minimum subsidence rate of 5 cm/yr at 9.3 kyr ago, followed by a progressive uplift of 10 m at a rate of approximately 9 mm/yr in relation to a magma reservoir pressure increase from 8.1 to 7 kyr ago. Then, slow subsidence caused by magma reservoir deflation and/or loading by new volcanic edifices occurred after 6.1 kyr. Finally, a new abrupt subsidence rate increase of more than 5 cm/yr started in 2018.

The abrupt subsidence rate that occurred 9.3 kyr ago is of the same order as the subsidence that occurred from May 2018 to January 2020 in Mayotte. The estimated downward surface deformation during the apex of the recent volcanic activity was 186 mm/yr, and the observed mean subsidence was 12 cm in 1.5 yr using the Global Navigation Satellite System (GNSS) and Interferometric Synthetic Aperture Radar (INSAR) during the 2018-2021 event [Lemoine et al., 2020].

A magma reservoir pressure decrease at 9.3 kyr B.P. or an increase from 8.1 kyr to 7.1 kyr is coherent with the volcanic context and explains the vertical movements. The observation of altered phonolite at 7.965 ± 45 kyr also suggests increased volcanic activity approximately 8 kyr ago [Zinke et al., 2003a]. Seismological findings show a complex geometry of a plume of hot upwelling rock rooted in the deep mantle, which is indicative of hotspot volcanoes beneath the Comoros archipelago [French and Romanowicz, 2015; O'Connor et al., 2019]. Geophysical observations are consistent with geochemical data showing a nonvolatile superplume isotopic signature in the Comoros–Mayotte and Madagascar systems [O'Connor et al., 2019]. The geochemical and petrographic characteristics of lavas from Mayotte indicate a deep mantle source [Spath et al., 1996] that suggests a complex hotspot origin for Mayotte volcanic activity [Emerick and Duncan, 1982; Debeuf, 2004]. The relatively long duration (1 kyr) of the deep overpressure (50-70 km) suggests that the magmatic pulse that has occurred during the past 10 kyr has been caused by deep mantle processes beneath Mayotte along the Comoros archipelago. This mantle/asthenospheric flow is potentially a pulse from a complex hotspot system.

The abrupt and relatively short (< 0.2 kyr) subsidence that occurred 9.3 kyr ago cannot be explained by the loading of new volcanic edifices due to the duration of isostatic adjustment (10 ± 5 kyr). Glacio-isostatic adjustment should be correlated with specific climatic change [Austermann et al., 2013], which is not the case here. Similarly, the significant uplift that occurred from 8.1 kyr to 7.1 kyr cannot be explained by abrupt mass unloading in this case. Landslides or erosion could cause isostatic adjustment, thereby influencing the subsidence or uplift rates and relative sea level curves, but over a longer duration [Gargani, 2022a and b]. These phenomena cannot explain an uplift with a duration of 1 kyr because the isostatic adjustment duration is approximately 10 times longer [Mitrovica et al., 2000; Van der Wal et al., 2010].

The estimated subsidence rate of 0.4 ± 0.1 mm/yr of Mayotte that occurred from 6.1 kyr to 1 kyr B.P. could have been caused by loading by new volcanic edifices and/or magma reservoir deflation. The slow subsidence of the island is considered active [Zincke et al., 2003; Camoin et al., 2004]. The estimated long-term subsidence rate is slightly higher than the long-term subsidence rate that is estimated to range from 0.13 mm/yr to 0.25 mm/yr over a longer duration [Camoin et al., 2004]. Neglecting vertical movement fluctuations of -10 m at 9.3 kyr B.P. followed by +10 m that occurred from 8.1 kyr B.P. to 7 kyr BP permits us to obtain an average subsidence value of approximately 2 m in 10 kyr B.P., which is in agreement with previous studies [Camoin et al., 2004].

This study suggests that new volcanic edifices were built between 6.1 kyr and 1 kyr ago. The volcanic activity during this period is indicated by relatively young volcanic morphological features east of Mayotte, along the slope of the main structure and by several volcanic rocks with ages ranging from 2.2 kyr to 7 kyr [Zinke et al., 2000, 2003; Camoin et al., 2004]. There is no published age concerning submarine edifices along the Mayotte slope to our knowledge.

## 5. Conclusion

During the last 10 kyr, significant volcanic activity caused uplift and subsidence in Mayotte. The subsidence of -10 m was due to a magma reservoir overpressure decrease at 9.3 kyr B.P., whereas the uplift of +10 m was caused by a magma reservoir pressure increase of 80 ± 20 MPa from 50-70 km deep. The significant duration of the overpressure was caused by a quasi-permanent deep magmatic flow during 1 kyr and suggests a deep magmatic root. The magma reservoir pressure increase was contemporaneous with the observed volcanic activity. The accumulation of volcanic material at Mayotte contributed to the slow subsidence of 2.5 m after 6.1 kyr B.P. The recent volcanic activity that started in 2018 could represent the last sequence of the deep-sourced cycle. The long volcanic cycle of approximately 10 kyr was recorded by relative sea level evolution and provides new insight for understanding the volcanic activity of the Comoros archipelago. Transient hotspot processes could have caused this long-term cycle.

**Acknowledgment**: